\documentclass[12pt]{article}
\usepackage{amssymb,epsf}
\def\lsim{\mathrel{\rlap {\raise.5ex\hbox{$ < $}}
{\lower.5ex\hbox{$\sim$}}}}
\def\gsim{\mathrel{\rlap {\raise.5ex\hbox{$ > $}}
{\lower.5ex\hbox{$\sim$}}}}
\def\sqr#1#2{{\vcenter{\vbox{\hrule height.#2pt
        \hbox{\vrule width.#2pt height#1pt \kern#1pt
           \vrule width.#2pt}
        \hrule height.#2pt}}}}

\def\lsim{{\displaystyle
{{\raise-8pt\hbox{$ <$}}
\atop{\raise5pt\hbox{$\sim$}}}}}
\def\gsim{{\displaystyle
{{\raise-8pt\hbox{$ >$}}
\atop{\raise5pt\hbox{$\sim$}}}}}
\def\slsim{{\displaystyle
{{\raise-8pt\hbox{$\scriptstyle <$}}
\atop{\raise5pt\hbox{$\scriptstyle \sim$}}}}}
\def\sgsim{{\displaystyle
{{\raise-8pt\hbox{$\scriptstyle  >$}}
\atop{\raise5pt\hbox{$\scriptstyle \sim$}}}}}
\newskip\humongous \humongous=0pt plus 1000pt minus 1000pt

\newcommand{\sumpf}[0]{\sum_{(H^{\rm f},G^{\rm f})}\! \! \! \!
{\raise
4pt
\hbox{$'$}}\,}

\newcommand{\sump}[0]{\sum_{(h,g)}\!{\raise 4pt \hbox{$'$}}\,}

\def\bs{\begin{subequations}}
\def\es{\end{subequations}}
\catcode`\@=11
\newcount\hour
\newcount\minute
\newtoks\amorpm
\hour=\time\divide\hour by60
\minute=\time{\multiply\hour by60 \global\advance\minute by-\hour}
\edef\standardtime{{\ifnum\hour<12 \global\amorpm={am}%
        \else\global\amorpm={pm}\advance\hour by-12 \fi
        \ifnum\hour=0 \hour=12 \fi
        \number\hour:\ifnum\minute<10 0\fi\number\minute\the\amorpm}}
\edef\militarytime{\number\hour:\ifnum\minute<10 0\fi\number\minute}
\def\draftlabel#1{{\@bsphack\if@filesw {\let\thepage\relax
   \xdef\@gtempa{\write\@auxout{\string
      \newlabel{#1}{{\@currentlabel}{\thepage}}}}}\@gtempa
   \if@nobreak \ifvmode\nobreak\fi\fi\fi\@esphack}
        \gdef\@eqnlabel{#1}}
\def\@eqnlabel{}
\def\@vacuum{}
\def\draftmarginnote#1{\marginpar{\raggedright\scriptsize\tt#1}}
\def\draft{\oddsidemargin -.2truein
        \def\@oddfoot{\sl preliminary draft \hfil
        \rm\thepage\hfil\sl\today\quad\militarytime}
        \let\@evenfoot\@oddfoot \overfullrule 3pt
        \let\label=\draftlabel
        \let\marginnote=\draftmarginnote
   \def\@eqnnum{(\theequation)\rlap{\kern\marginparsep\tt\@eqnlabel}%
\global\let\@eqnlabel\@vacuum}  }
\def\subequations{\refstepcounter{equation}%
  \edef\@savedequation{\the\c@equation}%
  \@stequation=\expandafter{\theequation}
  \edef\@savedtheequation{\the\@stequation}
  \edef\oldtheequation{\theequation}%
  \setcounter{equation}{0}%
  \def\theequation{\oldtheequation\alph{equation}}}

\def\endsubequations{\setcounter{equation}{\@savedequation}%
  \@stequation=\expandafter{\@savedtheequation}%
  \edef\theequation{\the\@stequation}\global\@ignoretrue
  \vspace*{-12pt} \\}
\def\bs{\begin{subequations}}
\def\es{\end{subequations}}
\relax
\def\Im{\,{\rm Im}\, }
\def\Re{\,{\rm Re}\, }

\def\g{\gamma}
\def\thefootnote{\fnsymbol{footnote}}
\def\be{\begin{equation}}
\def\ee{\end{equation}}
\def\ba{\begin{eqnarray}}
\def\ea{\end{eqnarray}}

\def\th{\vartheta}

\def\d{\delta}
\def\g{\gamma}

\def\sp{\ , \ \ }

\newcommand{\ar}[2]{{#1\atopwithdelims[]#2}}

\def\ee{\end{equation}}
\def\bea{\begin{eqnarray}}
\def\eea{\end{eqnarray}}
\def\nn{\nonumber}

\def\np#1#2#3{Nucl. Phys. {\bf{B#1}} (#2) #3}
\def\pl#1#2#3{Phys. Lett. {\bf{B#1}} (#2) #3}

\newcommand{\uarrw}[0]{\mathrel{
{\raise.5ex\vbox{\hrule width 1cm}\hskip-6pt\rightarrow}}}
\def\thebibliography#1{%
\vskip 0.5cm \centerline{\bf References}
\list{%
[\arabic{enumi}]}{\settowidth\labelwidth{[#1]}
\leftmargin\labelwidth
\advance\leftmargin\labelsep
\usecounter{enumi}}
\def\newblock{\hskip .11em plus .33em minus .07em}
\sloppy\clubpenalty4000\widowpenalty4000
\sfcode`\.=1000\relax}

\renewcommand{\theequation}{\arabic{section}.\arabic{equation}}

\renewcommand{\section}{\setcounter{equation}{0}\@startsection%
{section}{1}{0mm}{-\baselineskip}{0.5\baselineskip}%
{\normalfont\normalsize\bfseries}}

\renewcommand{\subsection}{\@startsection%
{subsection}{2}{0mm}{-\baselineskip}{0.5\baselineskip}%
{\normalfont\normalsize\slshape}}
\begin{document}
\renewcommand{\theequation}{\arabic{section}.\arabic{equation}}
\begin{titlepage}
\begin{flushright}
CERN-TH/99-108,\\ HUB-EP/99-20, \\
LPTENS/99-16,\\
hep-th/9904151 \\
\end{flushright}
\begin{centering}
\vspace{.15in}
\boldmath
{\bf \large
Four-Dimensional $N=2$ Superstring Constructions\\
\vskip .1cm
and their (Non-) Perturbative Duality Connections$^\ast$
}
\\
\unboldmath
\vspace{1.5 cm}
{Andrea GREGORI}$^{\ 1}$ and {Costas KOUNNAS}$^{\ 2}$\\
\medskip
\vspace{.3in}
{\it $^1 $ Humboldt-Universit\"{a}t, Institut f\"{u}r Physik,\\
D-10115 Berlin, Germany}\\
\medskip
{\it $^2 $ LPTH, \'{E}cole Normale Sup\'erieure, 24 rue Lhomond\\
F-75231, Paris Cedex 05, France\\
and\\
 Theory Division, CERN}
{\it 1211 Geneva 23, Switzerland}\\
\vspace{2cm}
{\bf Abstract}\\
\vspace{.1in}
We investigate the connections between four-dimensional, $N=2$
M-theory vacua constructed as orbifolds of type II, heterotic, and type I
strings. All these models have the same massless spectrum,
which contains an equal number of vector multiplets and hypermultiplets,
with a gauge group of the maximal rank allowed in a perturbative
heterotic string construction.
We find evidence for duality between two type I compactifications recently
proposed and a new heterotic construction that we present here.
This duality allows us to gain insight into the non-perturbative properties
of these models. In particular we consider gravitational corrections
to the effective action.
\end{centering}
\vspace{2cm}
\begin{flushleft}
CERN-TH/99-108\\
April 1999
\end{flushleft}
\hrule width 6.7cm
$^\ast$\  Research partially supported by the EEC under the contract\\
TMR-ERBFMRX-CT96-0045.\\
e-mail: gregori@physik.hu-berlin.de\\
$~~~~~~~~~~$
kounnas@nxth04.cern.ch

\end{titlepage}
\newpage
\setcounter{footnote}{0}
\renewcommand{\thefootnote}{\arabic{footnote}}

\setcounter{section}{1}
\section*{\normalsize{\bf 1. Introduction}}

In this work we consider five different string constructions,
which lead to the same low-energy spectrum. This
consists of a four-dimensional supergravity theory with
two space-time supersymmetries
whose spectrum, besides the gravity multiplet, contains
(when the gauge group is broken to its
Cartan subgroup) 19 vector multiplets and 20 hypermultiplets.
This massless spectrum is obtained via compactification and orbifold
projections from all the known perturbative string constructions, namely
the types IIA/B, the heterotic, and the type I.
The gauge group being a broken phase of
$SO(16) \times SO(16)$\footnote{When we speak of gauge group we mean,
on the heterotic string, the
part that comes from the currents. For the sake of
simplicity we always omit to mention the left and right gauge groups
which come from the compact space.},
this spectrum can be obtained from both the heterotic strings,
starting either from $E_8 \times E_8$ or from $SO(32)$.
In all these constructions the sypersymmetry and/or the gauge group
are broken by $Z_2$ projections.

The type II orbifolds are obtained
either with projections which act symmetrically on the right and left
movers \cite{vm,lc,gkr}, giving rise to type IIA or IIB models,
or with an asymmetric projection, in which all the
supersymmetries come only from the left movers.
These orbifolds are dual to one another: the type IIA and IIB
are trivially related by the inversion of an odd number of radii,
the compactification being self-mirror; the type II asymmetric orbifold
is instead related to these by a $U$-duality that exchanges
perturbative and non-perturbative moduli (see Ref.\cite{gkr}).

The type I orbifolds were recently constructed in Ref.\cite{adds}
as orientifolds of certain type IIB orbifolds, with a spontaneous
breaking of the $N=8$ supersymmetry.
In Ref.\cite{adds} they are indicated respectively as the
``Scherk--Schwarz breaking'' and the ``M-theory breaking'' models.

The heterotic orbifold, which we present here as a new construction,
is an interesting example of heterotic compactification in which,
although the gauge group has maximal rank, there are always, even
away from the Abelian point, an equal number of vector and hypermultiplets,
leading to vanishing gauge beta-functions.
It therefore behaves as the higher-level, reduced rank models
considered in Refs.\cite{fhsv,gkp}.
We provide evidence of duality between such heterotic orbifold
and the above two type I models, which
correspond to two different, but continuously related phases
of the heterotic theory. This allows us to see the
connection between the two models: their
relation is non-perturbative from the point of view of type I.
Thanks to the duality between these constructions,
we are able to determine at least part of the non-perturbative correction
to the effective coupling constant of a special, gravitational ($R^2$)
amplitude.
As a byproduct, we determine also the  $U$-duality group.
In particular, $S$-duality appears to be broken by the action of a freely
acting projection applied to the eleventh coordinate of M-theory.

The type IIA and heterotic constructions, on the other hand, are not dual.
This is related to the fact that the type IIA orbifold cannot be regarded
as a singular limit of a K3 fibration \cite{lc,gkr}.
However, even though these models cannot be compared for finite values
of the moduli, we argue that they correspond to two different
phases, or regions in the moduli space, of M-theory.
These two regions are connected in the moduli space.

The above issues are discussed according to the following order:

In Section 2 we review the type II orbifolds, which were discussed in detail
in Ref.\cite{gkr}. We discuss also the corrections to
the  $R^2$ term, which, being a function of the moduli of the vector
manifold,
will provide us with a quantity on which to test the duality relations.

In Section 3 we recall, from Ref.\cite{adds}, the type I orientifolds,
commenting  on the addition of discrete Wilson lines which break the
gauge group to the Cartan subgroup and discussing the gravitational
corrections.

In Section 4 we then discuss in detail the heterotic construction.
As the previous sections, also this ends with a discussion of the $R^2$
corrections.
Through the analysis of this term, we discuss also the duality
relation between this model and the type I of Section 3.
The duality relations are used to obtain (at least part of) the
non-perturbative gravitational corrections, as well as an insight into
other non-perturbative properties.

Finally, in Section 5 we comment on the connections between the
various phases of these $N=2$, M-theory compactifications.
Our conclusions are given in Section 6.

\noindent

\vskip 0.3cm
\setcounter{section}{2}
\setcounter{equation}{0}
\section*{\normalsize{\bf 2. The type II constructions}}

\subsection*{\normalsize{\sl 2.1. The type IIA construction}}

We start by reviewing the construction of the type IIA,
which is obtained by compactification
of the ten-dimensional superstring on a Calabi--Yau manifold
with Hodge numbers $h^{1,1}=h^{2,1}=19$ \cite{vm,lc}.
We recall that
this manifold has an orbifold limit in which the $N=8$ supersymmetry
of the type IIA string compactified on $T^6$
is reduced to $N=2$ by two $Z_2$ projections,
$Z_2^{(1)}$ and $Z_2^{(2)}$ \cite{lc}.
The action of these on $T^6=T^2_{(1)} \times T^2_{(2)} \times T^2_{(3)}$
is the following:
$Z_2^{(1)}$ acts as a rotation in $T^2_{(1)} \times T^2_{(3)}$,
while $Z_2^{(2)}$ acts as a rotation on $T^4=T^2_{(2)} \times T^2_{(3)}$
and as a translation in $T^2_{(1)}$.
The partition function of this orbifold reads (see Ref.\cite{gkr}):
\ba
Z_{\rm II}^{(1,1)} & = &
{1 \over \Im \tau \vert \eta \vert^{24} } {1 \over 4}
\sum_{H^1,G^1} \sum_{H^2,G^2}
\Gamma_{6,6} \ar{H^1,H^2}{G^1,G^2} \nonumber \\
&&\nonumber \\
&& \times {1 \over 2} \sum_{a,b} (-)^{a+b+ab} \vartheta \ar{a}{b}
\vartheta \ar{a+H^1}{b+G^1}
\vartheta \ar{a+H^2}{b+G^2}
\vartheta \ar{a-H^1-H^2}{b-G^1-G^2} \nonumber \\
&&\nonumber \\
&& \times {1 \over 2} \sum_{\bar{a},\bar{b}}
(-)^{\bar{a}+\bar{b}+\bar{a}\bar{b}} \bar{\vartheta}
\ar{\bar{a}}{\bar{b}}
\bar{\vartheta} \ar{\bar{a}+H^1}{\bar{b}+G^1}
\bar{\vartheta} \ar{\bar{a}+H^2}{\bar{b}+G^2}
\bar{\vartheta} \ar{\bar{a}-H^1-H^2}{\bar{b}-G^1-G^2}\, ,
\label{zII}
\ea
where
the contribution of  the compactified bosons,
$X^I, \bar X^I,I=1,\ldots,6$ is contained in the factor
$\Gamma_{6,6} \ar{H^1,H^2}{G^1,G^2} \Big /
\vert \eta \vert^{12}$, while
the other factors contain the contribution of
their fermionic superpartners, $\Psi^I$, ${\bar\Psi}^I$, of the
left- and right-moving  non-compact supercoordinates
$X^{\mu}$, $\Psi^{\mu}$, ${\bar X}^{\mu}$, ${\bar \Psi}^{\mu}$
and of the super-reparametrization ghosts
$b,c,\beta,\gamma$ and ${\bar b},{\bar c},{\bar \beta},{\bar
\gamma}$.
$(H^{1},G^{1})$ refer to the boundary conditions
introduced by the projection $Z_2^{(1)}$, and $(H^{2},G^{2})$ to the
projection $Z_2^{(2)}$.

$\Gamma_{6,6} \ar{H^1,H^2}{G^1,G^2}$
factorizes into the contributions corresponding to the three
tori of $T^6$:
\be
\Gamma_{6,6} \ar{H^1,H^2}{G^1,G^2}=
\Gamma_{2,2}^{(1)} \ar{H^1 \vert H^2}{G^1 \vert G^2}\,
\Gamma_{2,2}^{(2)} \ar{H^2 \vert 0}{G^2 \vert 0}\,
\Gamma_{2,2}^{(3)} \ar{H^1+H^2 \vert 0}
{G^1+G^2 \vert 0}\, ,
\ee
which are expressed in terms of the twisted and shifted
characters of a $c=(2,2)$ block,
$\Gamma_{2,2} \ar{h\vert h'}{g\vert g'}$; the first column refers to
the twist, the second to the shift. The non-vanishing components
are the following:
\ba
\Gamma_{2,2} \ar{h\vert h'}{g\vert g'}
&=&
{4\,  \vert\eta \vert^6\over
\left\vert
\vartheta{1+h\atopwithdelims[] 1+g}\,
\vartheta{1-h\atopwithdelims[] 1-g}
\right\vert} \sp
{\rm for\ } (h',g')=(0,0) \
{\rm or\ } (h',g')=(h,g) \nn \\
&=&
\Gamma_{2,2} \ar{h'}{g'} \sp ~~~~~~~~~
{\rm for\ } (h,g)=(0,0)\, ,
\label{g22ts}
\ea
where $\Gamma_{2,2} \ar{h'}{g'}$ is the $Z_2$-shifted $(2,2)$ lattice
sum. In this case, the only shift is that due to the $Z_2^{(2)}$
translation on $T^2_{(1)}$, $(-)^{m_2 G^2}$.

For a detailed analysis of the spectrum of this model, we refer
to Ref.\cite{gkr}. Here we simply recall that the $Z_2^{(1)}$ twisted sector
has sixteen fixed points, which give rise to eight vector
and eight hypermultiplets. The twist of $Z_2^{(2)}$ on the other hand is
accompanied by a lattice shift, so there are no massless states from this
twisted sector. The other eight vector and eight hypermultiplets come
from the sector twisted by $Z_2^{(1)} \times Z_2^{(2)}$
(the $H^{(1)}+H^{(2)}$-twisted sector).

We now consider the corrections
to the $R^2$ term. They receive a non-zero contribution at one loop,
and are related to the infrared-regularized integral of the fourth
helicity supertrace, $B_4$ (for the definition and the details
we refer, for instance, to Ref.\cite{gkr}).
The one-loop correction to the coupling constant is:
\ba
{16 \, \pi^2 \over g^2_{\rm grav} \left( \mu^{(\rm IIA)} \right)}
& = & -2 \log \Im T^1 \vert \th_4 \left( T^1  \right)  \vert^4
- 6 \log \Im T^2 \vert \eta \left( T^2 \right)  \vert^4
- 6 \log \Im T^3 \vert \eta \left( T^3 \right)  \vert^4 \nn \\
&& + 14 \log {M^{(\rm IIA)} \over \mu^{(\rm IIA)}} \, ,
\label{IIthr}
\ea
where $T^1,T^2,T^3$ are the K\"{a}hler class moduli of the three tori of the
compact space.
The last term in Eq.(\ref{IIthr}),
which encodes the infrared running due to the massless
contributions, is expressed in terms of the type IIA
string scale $M^{(\rm IIA)} \equiv {1 / \sqrt{\alpha'_{\rm IIA}}}$
and infrared cut-off $\mu^{(\rm IIA)}$.
The coefficient, 14, is actually the massless contribution to
$(B_4-B_2) \big/ 3$, as is computed in field theory (see Ref.\cite{gkp}).
Observe that there is no limit in the space of moduli
$T^1$, $T^2$, $T^3$, at which this correction reproduces the behaviour of
an $N=4$ orbifold, for which it is expected to depend
on the K\"{a}hler class modulus of only one torus
(see for instance Refs.\cite{hmn=4,6auth}).
This implies that, in the space of these three moduli, there
is no region in which the $N=4$ supersymmetry is restored, as
happens instead in orbifold constructions with a spontaneous breaking
of supersymmetry, such as those considered in Refs.\cite{gkp,gkp2}.
There is therefore no perturbative connection to an $N=4$ theory,
and therefore this orbifold cannot be seen as a singular limit in the
moduli space of a K3 fibration \cite{gkr}.

As explained in Ref.\cite{gkr},
the type IIB dual compactification is trivially obtained
by changing the chirality of the right-moving spinors.
This is obtained by changing the phase
$(-)^{\bar{a}+\bar{b}+\bar{a}\bar{b}}$
in Eq.(\ref{zII}) to $(-)^{\bar{a}+\bar{b}}$.
The analysis is similar and we obtain analogous results,
with the role of the fields $T^i$, $i=1,2,3$,
associated to the K\"{a}hler classes of the three tori, interchanged with
that of the fields $U^i$, associated to the complex structures.

\subsection*{\normalsize{\sl 2.2. The type II asymmetric dual}}

The model above described possesses a type II dual constructed
as an asymmetric orbifold \cite{gkr},
obtained by combining the projection $Z_2^{(1)}$,
acting in the same way as before, with $Z_2^{\rm F}$, which projects out
all the right-moving supersymmetries by relating the action of
the right fermion number operator, $(-)^{\rm F_R}$,
to a translation in the compact space.
The partition function reads \cite{gkr}:
\ba
Z_{\rm II}^{(2,0)} & =  & {1 \over \Im \tau |\eta|^{24} }
{1 \over 4} \sum_{H^{\rm F},G^{\rm F}} \sum_{H^{\rm o},G^{\rm o}}
\Gamma_{6,6} \ar{H^{\rm F}, H^{\rm o}}{G^{\rm F}, G^{\rm o}} \nonumber \\
&& \times {1 \over 2}
\sum_{a,b}(-)^{a+b+ab}
\vartheta^2 \ar{a}{b}
\vartheta \ar{a+H^{\rm o}}{b+G^{\rm o}}\vartheta \ar{a-H^{\rm
o}}{b-G^{\rm o}}
 \nonumber \\
&& \times {1 \over 2}
\sum_{\bar{a},\bar{b}}(-)^{\bar{a}+\bar{b}+\bar{a}\bar{b}}
(-)^{\bar{a}G^{\rm F}+\bar{b}H^{\rm F}+H^{\rm F}G^{\rm F}}
\bar{\vartheta}^2 \ar{\bar{a}}{\bar{b}}
\bar{\vartheta} \ar{\bar{a}+H^{\rm o}}{\bar{b}+G^{\rm o}}
\bar{\vartheta} \ar{\bar{a}-H^{\rm o}}{\bar{b}-G^{\rm o}}~,
\label{z2as}
\ea
with
\be
\Gamma_{6,6} \ar{H^{\rm F}, H^{\rm o}}{G^{\rm F}, G^{\rm o}} =
\Gamma_{2,2}^{(1)} \ar{H^{\rm o}~|~H^{\rm F}}{G^{\rm o}~|~G^{\rm F}}
\Gamma_{2,2}^{(2)} \ar{H^{\rm o}~|~0}{G^{\rm o}~|~0}
\Gamma_{2,2}^{(3)} \ar{0~|~0}{0~|~0}~.
\label{z20}
\ee
As was discussed in Ref.\cite{gkr}, the $R^2$ gravitational correction,
which receives contribution only from one loop, is
a function of the moduli $T^{\rm As}$, $U^{\rm As}$, associated
respectively to the K\"{a}hler class and the complex structure of the third
complex plane; it reads
\ba
{16 \, \pi ^2 \over g^2_{\rm grav} \left( \mu^{(\rm As)} \right)} & = &
-6 \log \Im T^{\rm As} \left\vert \eta \left( T^{\rm As} \right)
\right\vert^4\, -
6 \log \Im U^{\rm As} \left\vert \eta \left( U^{\rm As} \right)
\right\vert^4 \, + \nn \\
&& +  14 \log {M^{(\rm As )} \over \mu^{(\rm As )}}~,
\label{dAs1}
\ea
where we introduced the type II asymmetric mass scale and
infrared cut-off, $M^{(\rm As )}$ and $\mu^{(\rm As )}$ respectively.
A comparison of Eq.(\ref{dAs1}) and Eq.(\ref{IIthr}) leads directly to the
identifications $T^2 = T^{\rm As}$, $T^3 = U^{\rm As}$
and $T^1=\tau^{\rm As}_S$, where $\tau^{\rm As}_S \equiv 4 \pi S^{\rm As}$,
$S^{\rm As}$ being the dilaton--axion field of the type II asymmetric
models\footnote{We recall that this field belongs to a vector multiplet,
as in the heterotic constructions \cite{gkr}.}.
For later use, we note here that the corrections
Eq.(\ref{IIthr}) and Eq.(\ref{dAs1}) remain the same if
we correct the  $R^2$ term by adding gauge amplitudes $F^2$:
owing to the absence, in the perturbative type II strings,
of gauge charges, these amplitudes are identically vanishing.

\noindent

\vskip 0.3cm
\setcounter{section}{3}
\setcounter{equation}{0}
\section*{\normalsize{\bf 3. The type I models}}

There are two type I orbifolds that possess the desired massless spectrum.
They were constructed in Ref.\cite{adds}, as orientifolds
of type IIB orbifolds, with the $N=8$ supersymmetry
spontaneously broken to $N=4$ by a freely acting projection,
$Z_2^{(\rm f)}$. The latter
acts as a twist on $T^4$, and, in the first case,
as a translation in the momenta,
produced by the projection $(-)^{m G^{\rm f}}$,
on a circle of $T^2$. In Ref.\cite{adds} this construction is called the
``Scherk--Schwarz breaking'' model; in the following we will refer to it as
the model {\bf A}. Model {\bf B},
referred to in Ref.\cite{adds} as the ``M-theory breaking'' model,
is obtained when the translation is performed on the windings instead.
After the orientifold projection, the resulting type I models
possess an $N=4$ supersymmetry spontaneously broken to $N=2$.
In model {\bf A} the $N=4$ is restored
in the limit of decompactification (large radius)
of the circle translated by $Z_2^{(\rm f)}$; in model {\bf B},
instead, it is restored when the radius goes to zero.
Because of the spontaneous nature of the breaking of the $N=4$
supersymmetry, the massless spectrum contains an equal
number of vector and hypermultiplets\footnote{As usual, we
don't count here the three vector multiplets and the four
hypermultiplets originating from the compact space, which are common
to all the
$N=2$ orbifolds we consider in this paper.} in the two models.
However, while in model {\bf A} the gauge group consists only of
factors arising from the contribution of
99-brane sector states, in model {\bf B} the consistency conditions
of the construction, as derived by imposing tadpole cancellation,
require the presence of both D9- and D5-branes, and the
gauge group is the product of the 99-brane sector and
the 55-brane sector contribution.

Model {\bf A}, in which the gauge group
is a broken phase of $SO(32)$,
is expected to possess a heterotic dual, and indeed we will
construct such a dual in the next section.
As explained in \cite{adds}, there is a wide choice of
Wilson lines compatible with this Scherk--Schwarz projection, leading
to different breakings of the gauge group.
In order to compare the type I and the heterotic constructions,
we introduce  a further set of discrete Wilson lines, which
break the gauge group to its Cartan subgroup, $U(1)^{16,}$
\footnote{Here also we omit the contribution coming from the compact
space.}.
In this way we obtain $3+N_V$ vector multiplets and $4+N_H$ hypermultiplets,
with $N_V=N_H=16$. It is possible to choose the Wilson lines to act
as $Z_2$ shifts in the directions twisted by $Z_2^{(\rm f)}$.
This choice corresponds to a breaking of the gauge group
at the $N=2$, six-dimensional level, before the further torus
compactification and supersymmetry breaking via orbifold projection;
these Wilson lines therefore do not enter the $\Gamma_{2,18}$
lattice explicitly.
The only vector multiplets moduli that appear are those of the
$\Gamma_{2,2}$ lattice associated with the two-torus of the compactification
from six to four dimensions.
We will test the duality with the heterotic construction of the next
section,
through a comparison of the ``holomorphic'' gravitational corrections.
These are defined as the corrections to the effective
coupling constant of a special combination of $R^2$ and $F^2$
terms \cite{gkp,gkp2}, namely
\be
\langle R^{\prime 2}_{\rm grav} \rangle \equiv
\langle R^{2}_{\rm grav} \rangle +
{1\over 12} \langle F_{\mu \nu} F^{\mu \nu} \rangle_{T^2}  +
{5 \over 48} \langle F_{\mu \nu} F^{\mu \nu} \rangle_{\rm gauge} \, .
\label{r2f2}
\ee
This combination was introduced in the above cited
works in the context of heterotic orbifolds; it was shown to possess
an amplitude smooth in the moduli space, in which the non-harmonic
contributions to the $R^2$ and the $F^2$ amplitudes
cancel each other, leaving only the
$\Gamma_{2,2}$ ($Z_2$-shifted) lattice sum. This is
due to the presence of the full bunch of states, contained in the
part of the heterotic spectrum which has, as massless excitations,
the states
of the $c=(0,16)$ currents.
On the type I side, the pure $R^2$ amplitude, although smooth
in the moduli space,
contains, besides the lattice sum provided by the torus ${\cal T}$,
the contributions of the Klein bottle ${\cal K}$,
the annulus $\cal{A}$ and the M\"{o}bius strip
$\cal{M}$ \cite{ms}--\cite{apt}.
However, only $N=2$ BPS multiples contribute to such amplitudes,
which therefore are all proportional to a ``supersymmetric index''
\cite{ms}; for the combination of gravitational
and gauge amplitudes Eq.(\ref{r2f2}),
the $\cal{K}$, $\cal{A}$ and $\cal{M}$ contributions cancel
(notice that in the case of type I, the gauge amplitude
$\langle F_{\mu \nu} F^{\mu \nu}   \rangle_{T^2}$
vanishes, because these states come from the RR sector of the
type IIB string).
The only contribution therefore comes from the torus ${\cal T}$.

Since in this model there are no D5-branes,
we expect the tree-level effective coupling ${1 \over g^2}$
to be given by the imaginary part of only
one complex field, $S=S_1+iS_2$, whose real part $S_1$
is the scalar dual to $B_{\mu \nu}$, while the imaginary part is
\be
S_2={\rm e}^{-\phi_4} G^{1/4} \omega^2~,
\ee
where $\phi_4$ is the dilaton of the four-dimensional
compactification, $\sqrt{G} \sim R_4 R_5$ is the volume of the
two-torus, and
$\omega^4$ is the volume of the K3, which in the case at hand is in his
$T^4 \big/ Z_2$ limit.
This situation has to be contrasted with the most general one
\cite{abfpt,sgn}, in which the coupling is given by a combination
\be
vS_2+v'S_2'
\label{vsvsp}
\ee
in which $S_2'={\rm e}^{-\phi_4} G^{1/4} \omega^{-2}$
is part of a complex field $S'$ whose real part $S_1'$ is the dual
of $B_{45}$. The part of the coupling proportional to the inverse of the
K3 volume is due to the presence of D5-branes.

The one-loop contribution is given by the complex structure-dependent
part of the integral, over the fundamental domain,
of the $\Gamma_{2,2}$, $Z_2$-shifted lattice sum appearing in the ${\cal T}$
amplitude. This can be computed in an
infrared-regularized background, as in Ref.\cite{infra}. This would lead to
the introduction of a curvature in the space-time, providing a
cut-off $\mu$ that, once the flat limit is taken,
appears in the running of the effective
coupling \cite{infra}--\cite{gauge}.
There is no need to go into the details of such a
procedure: such a running is in fact fixed by a regularization prescription,
which imposes the matching of field theory and string computations
in the infrared, and can therefore be determined simply by field-theory
arguments. On the other hand, the full dependence of the corrections on the
modulus $U$, the complex structure of the torus $T^2$ and the only
non-trivial modulus that appears in such terms, the
Wilson lines being frozen to fixed values,
can be easily derived by knowing the  action of $Z_2^{(\rm f)}$ on the
torus \cite{kkprn}.
The total result, including tree-level and one-loop contributions,
is therefore given by
\be
{16 \, \pi^2 \over g^2_{\rm   grav}(\mu^{(\rm I)})}  =
16 \, \pi^2 \Im S
-2 \log \Im U \left \vert \vartheta_4 \left( U \right) \right\vert^4
+14 \log {M^{(\rm I )} \over \mu^{(\rm I )}}~,
\label{Ithr}
\ee
where the Jacobi function $\vartheta_4$ corresponds to a translation
$(-)^{m_2 G^{\rm f}}$, in the second circle of $T^2$.
The $SL(2,Z)_U$ duality group is broken to a $\Gamma(2)$
subgroup (see Refs.\cite{gkp,6auth,gkp2,kkprn}).
In the last term we collect
the dependence on the infrared cut-off $\mu^{(\rm I)}$ and the type I
string mass scale $M^{(\rm I)}$.

We now consider the ``M-theory breaking'', model {\bf B}.
We specialize to the case in which, with reference to the notation of
Ref.\cite{adds}, $n_2=d_2=0$, i.e. the symplectic
factors do not appear in the gauge group, which is therefore given by
$SO(16)_{99} \times SO(16)_{55}$ (the subscripts indicate the origin
of these factors).
We now introduce Wilson lines, as we did in the previous case,
to break $SO(16)_{99}$ to $U(1)^8$.
We also move the D5-branes a bit far from each other, in order to break
also the second factor to the Cartan subgroup.
We can now repeat the same arguments as before and compute
the analogous ``gravitational'' corrections.
In this case, due to the presence of the D5-branes, we expect a dependence
of the tree level effective coupling also on the field $S'$.
We therefore obtain:
\be
{16 \, \pi^2 \over g^2_{\rm   grav}(\mu^{(\rm I)})}  =  16 \, \pi^2
v \Im S +16 \, \pi^2 v' \Im S'
-2 \log \Im U \left \vert \vartheta_4 \left( U \right) \right\vert^4
+14 \log {M^{(\rm I )} \over \mu^{(\rm I )}}~,
\label{Ithrp}
\ee
where, as in Refs.\cite{abfpt,apt},
we allow for the presence of two independent
contributions to the effective coupling constant ($v v' \neq 0$).
Actually, since the model is symmetric under the exchange of
the D9- and the D5-branes sectors, we deduce that $v=v'=1$.
The $\vartheta_4(U)$ is obtained for a translation
$(-)^{n_1 G^{\rm f}}$ on the windings of the first circle of $T^2$.

\noindent

\vskip 0.3cm
\setcounter{section}{3}
\setcounter{equation}{0}
\section*{\normalsize{\bf 3. The heterotic construction}}

We now discuss in detail the heterotic dual.
This is constructed as a $Z_2$ freely acting orbifold of the
heterotic string compactified on $T^6$, with the
gauge group broken to $SO(16)\times SO(16)$
by a $Z_2$, discrete Wilson line.
The type I construction corresponds to a special region in the
moduli space of the theory, and in order to be able to
compare heterotic and type I,
we must choose on the heterotic side a special set of Wilson lines.
What we need is a set of three other discrete Wilson lines, which act
on the compact space as $Z_2$ translations, and further break
$SO(16) \times SO(16)$ to $U(1)^{8} \times SO(4)^4$.
The $N=4$ supersymmetry is then reduced to $N=2$
by a further $Z_2$ freely acting projection, $Z_2^{(\rm f)}$,
which moreover acts as a further Wilson line, which breaks
the gauge group, leaving massless only the bosons in the Cartan
subgroup.
The partition function, $Z_{\rm Het}$, can be easily written in terms
of the usual fermionic and bosonic characters for the compact space
and the $c=(0,2)$, $SO(4)$ twisted  characters
introduced in Ref.\cite{gkp}, $F_1$ and $F_2$, for the $c=(0,16)$
currents. We recall that:
\be
F_{1}\ar{\g,h}{\d, g} \equiv {1\over {\eta}^2}\,
{\vartheta}^{1/2}
\ar{\g+h_1}{\d+g_1}\,
{\vartheta}^{1/2}
\ar{\g+h_2}{\d+g_2}\,
{\vartheta}^{1/2}
\ar{\g+h_3}{\d+g_3}\,
{\vartheta}^{1/2}
\ar{\g-h_1-h_2-h_3}{\d-g_1-g_2-g_3}
\ee
and
\be
F_{2}\ar{\g, h}{\d, g} \equiv {1\over {\eta}^2}\,
{\vartheta}^{1/2}
\ar{\g}{\d}\,
{\vartheta}^{1/2}
\ar{\g+h_1-h_2}{\d+g_1-g_2}\,
{\vartheta}^{1/2}
\ar{\g+h_2-h_3}{\d+g_2-g_3}\,
{\vartheta}^{1/2}
\ar{\g+h_3-h_1}{\d+g_3-g_1}\, ,
\ee
where $h\equiv (h_1,h_2,h_3)$ and similarly for $g$.
Under $\tau\to \tau+1$, $F_{I}$ transforms as:
\ba
F_{1}\ar{\g, h}{\d, g} & \to &
F_{1}\ar{\g, h}{\g+\d +1, h + g} \nn \\
&& \times
\exp -{i\pi\over 4}
\left({2\over 3} - 4\g +2 \gamma^2+h_1^2+h_2^2 +h_3^2 + h_1 h_2+h_2
h_3+h_3 h_1 \right) \, ,
\\
F_{2}\ar{\g, h}{\d, g} & \to &
F_{2}\ar{\g, h}{\g+\d +1, h + g} \nn \\
&& \times
\exp -{i\pi\over 4}
\left({2\over 3} - 4\g +2 \gamma^2+h_1^2+h_2^2 +h_3^2 - h_1 h_2-h_2
h_3-h_3 h_1 \right)\, .
\ea
In terms of these, we have
\ba
Z_{\rm Het} & = &
{1 \over \Im \tau   | \eta|^4 } {1 \over 2}
\sum_{H^{\rm f},G^{\rm f}}
Z_{6,22} \ar{H^{\rm f}}{G^{\rm f}} \nonumber \\
&& \times {1 \over 2} \sum_{a,b}{1\over \eta^4}~ (-)^{a+b+ab}
\vartheta \ar{a}{b}^2
\vartheta \ar{a+H^{\rm f}}{b+G^{\rm f}}
\vartheta \ar{a-H^{\rm f}}{b-G^{\rm f}} \, ,
\label{hH}
\ea
where the second line stands for the contribution of the 10 left-moving
world-sheet fermions $\psi^{\mu},\Psi^I$ and the ghosts
$\beta,\gamma$ of the super-reparametrization;
$Z_{6,22}\ar{H^{\rm f}}{G^{\rm f}}$ accounts for the $(6,6)$
compactified coordinates and the $c=(0,16 )$ conformal system, which
is described by the 32 right-moving fermions $\Psi_A$, $A=1,\ldots,32$:
\be
Z_{6,22}\ar{H^{\rm f}}{G^{\rm f}}=
{1\over 2^{5}}\, \sum_{\vec h, \vec g}~{1\over \eta^6 {\bar \eta}^6 }
\Gamma_{2,2} \ar{H^{\rm f},h_1}{G^{\rm f},g_1}
\,
\Gamma_{4,4} \ar{H^{\rm f}\vert \vec h}{G^{\rm f}\vert \vec g}\,
{\overline \Phi}\ar{H^{\rm f}, \vec h}{G^{\rm f}, \vec g}\,
,\label{hH622}
\ee
and
\ba
\Phi\ar{H^{\rm f}, \vec h}{G^{\rm f}, \vec g}
& = & {1 \over 2} \sum_{\g,\d}
F^2_1 \ar{\g,h_1,h_2,h_3}{\d,g_1,g_2,g_3}
_{(H^{\rm f},G^{\rm f})}
F^2_2 \ar{\g,h_1,h_2,h_3}{\d,g_1,g_2,g_3}
_{(H^{\rm f},G^{\rm f})} \nn \\
&& \times \,
F^2_1 \ar{\g+h_4,H^{\rm f},h_2,h_3}{\d+g_4,G^{\rm f},g_2,g_3}
F^2_2 \ar{\g+h_4,H^{\rm f},h_2,h_3}{\d+g_4,G^{\rm f},g_2,g_3}~.
\label{phi}
\ea
In Eq.(\ref{hH622}) we used the twisted-shifted bosonic character
for $\Gamma_{4,4} \ar{h|h'}{g|g'}$, in which the first column indicates
the twist $(h,g)$, the second the shift $(h',g')$, and the
doubly-shifted character
of the two-torus $\Gamma_{2,2} \ar{h,h'}{g,g'}$.
In Eq.(\ref{phi}) the subscripts $(H^{\rm f},G^{\rm f})$ indicate the  
embedding
of the spin connection in the gauge group. This is realized explicitly
through a modification of the arguments in the first Ising character,
both in $F_1$ and in $F_2$:
\[
F_{1}\ar{\g,h}{\d, g}_{(H^{\rm f},G^{\rm f})} \equiv {1\over {\eta}^2}\,
{\vartheta}^{1/2}
\ar{\g+h_1+H^{\rm f}}{\d+g_1+G^{\rm f}}\,
{\vartheta}^{1/2}
\ar{\g+h_2}{\d+g_2}\,
{\vartheta}^{1/2}
\ar{\g+h_3}{\d+g_3}\,
{\vartheta}^{1/2}
\ar{\g-h_1-h_2-h_3}{\d-g_1-g_2-g_3}~;
\]
\be
F_{2}\ar{\g, h}{\d, g}_{(H^{\rm f},G^{\rm f})} \equiv {1\over {\eta}^2}\,
{\vartheta}^{1/2}
\ar{\g+H^{\rm f}}{\d+G^{\rm f}}\,
{\vartheta}^{1/2}
\ar{\g+h_1-h_2}{\d+g_1-g_2}\,
{\vartheta}^{1/2}
\ar{\g+h_2-h_3}{\d+g_2-g_3}\,
{\vartheta}^{1/2}
\ar{\g+h_3-h_1}{\d+g_3-g_1}\, ,
\ee
With this embedding, the
shift in $\Gamma_{2,2}$ is produced by the projection
$(-)^{m_2G^{\rm f}+n_2g_1}$ \footnote{As in Refs.\cite{gkp,gkp2},
there exists also an alternative heterotic
construction with the same massless spectrum,
in which the spin connection is embedded in the right $U(1)^2$ factor
from the untwisted two-torus. In this case, the
shift in $\Gamma_{2,2}$ is produced by an asymmetric projection, e.g.
$(-)^{(m_1+n_1)G^{\rm f}}$. We will not consider this alternative
construction, because it is not dual to the type I.}.
The conformal blocks in the second line of Eq.(\ref{phi}) provide
the right-moving part of eight
vector multiplets, corresponding to a factor $U(1)^8$ in the gauge group,
and sixteen hypermultiplets. From the blocks on the r.h.s. of the first line
we get the other eight vectors, to make $U(1)^{16}$, and no
further hypermultiplets. In addition to these, the massless
spectrum of the model contains the usual three vector multiplets
and four hypermultiplets of an ordinary $T^2 \times T^4 / Z_2$
compactification of the heterotic string.
Thanks to the free action of all the projections, there are no additional
massless states coming from the twisted sectors.
It is worth remarking that, when some of the Wilson lines are absent,
the gauge group is enlarged but still $N_V=N_H$.
Notice that, because of the embedding of the spin connection into
the gauge group, it is not possible to construct this model at the point
$SO(16) \times SO(16)$, but only at a broken phase of it.

It is easy to recognize that the $N=2$ sector of this orbifold,
specified by $(H^{\rm f},G^{\rm f}) \neq (0,0)$, belongs to the same
universality class as the $N=2$
heterotic constructions with $N_V=N_H$, considered in Refs.\cite{gkp,gkp2}.
In fact, the modular transformation properties of the
untwisted $\Gamma_{2,2}$ lattice, toghether with the
condition $N_V=N_H$, as was already pointed out in Ref.\cite{gkp}, are
sufficient to fix this orbifold sector uniquely, the difference
between the various models residing in the $N=4$ sector.
A consequence of this is that not only the $N=2$ singularities,
but also all the threshold corrections that receive contribution only from
this sector, are the same as for the other models of this class;
we therefore skip the details about the analysis of this
model and go directly to the discussion of the
gravitational corrections.

As we anticipated in Section 2, it was pointed out in Refs.\cite{gkp,gkp2}
that the ``pure'' gravitational
amplitude, $\langle R^2 \rangle$, must be corrected with a term
proportional to the gauge amplitude,
$\langle F_{\mu \nu} F^{\mu \nu} \rangle$, in order to make it
holomorphic and non-singular.
We recall here that the precise combination is:
\be
\langle R^{\prime 2}_{\rm grav} \rangle \equiv
\langle R^{2}_{\rm grav} \rangle +
{1\over 12} \langle F_{\mu \nu} F^{\mu \nu} \rangle_{T^2}  +
{5 \over 48} \langle F_{\mu \nu} F^{\mu \nu} \rangle_{\rm gauge} \, .
\label{r2f2h}
\ee
The tree-level plus one-loop contribution reads:
\ba
{16 \, \pi^2 \over g^2_{\rm   grav}(\mu^{(\rm Het)})}
& = & 16 \, \pi^2 \Im S^{(\rm Het)}
-2 \log \Im T \left \vert \vartheta_4 \left( T \right) \right\vert^4
-2 \log \Im U \left \vert \vartheta_4 \left( U \right) \right\vert^4
\nn \\
&& +14 \log {M^{(\rm Het)} \over \mu^{(\rm Het)} } + \ {\rm const.}
\, ,
\label{htr}
\ea
where $S^{(\rm Het)}$ is the heterotic axion--dilaton field,
\be
\Im S^{(\rm Het)}  ={1 \over g^2_{\rm Het} }
\ee
and we used the string scale $M^{(\rm Het)} \equiv
{1 / \sqrt{\alpha'_{\rm Het}}}$ and the infrared cut-off
$\mu^{(\rm Het)}$ of the heterotic string.

Owing to the free action of the projection $Z_2^{(\rm f)}$,
also in this model the $N=4$ supersymmetry
is spontaneously broken perturbatively \cite{kk}; it is restored when
$T \to \infty$, $U \to \infty$. This limit corresponds, as in the type I
construction, to a decompactification to five dimensions
\footnote{When $\Re T=\Re U=0$, we have
$\Im T \sim R_1 R_2$ and $\Im U \sim R_2/R_1$, where $R_1$, $R_2$ are
the radii of the two circles of the torus. In this case
this limit corresponds to $R_2 \to \infty$, with $R_1$ fixed.}.
In order to compare this model with the type I, ``Scherk--Schwarz
breaking'' model {\bf A} of the previous section,
we consider the limit in which $T$ is large while
$U$ is kept finite (we stress that in this limit the supersymmetry
remains broken to $N=2$).
In this limit, the correction Eq.(\ref{htr}) depends on $T$ only
logarithmically:
\be
{16 \, \pi^2 \over g^2_{\rm   grav}(\mu^{(\rm Het)})}
 \approx  16 \, \pi^2 \Im S^{(\rm Het)}
-2 \log \Im U \left \vert \vartheta_4 \left( U \right) \right\vert^4
+ {\cal O}(\log \Im T)~.
\label{htr1}
\ee
If we discard this logarithmic dependence for the moment, we see that
Eq.(\ref{htr1}) exactly reproduces the type I correction given in
Eq.(\ref{Ithr}), with the identifications $S^{(\rm Het)} \equiv S$
and $U^{(\rm Het)} \equiv U^{(\rm I)}$.
We can interpret the logarithmic dependence as due to effects that
are non-perturbative from the type I point of
view\footnote{This is the same phenomenon as that appearing in the examples
of type IIA/type II asymmetric orbifolds dualities considered
in Refs.\cite{gkr,6auth,gkp2}, where the absence of tree level dilaton
dependence
in the analogous corrections on the asymmetric orbifolds indeed corresponds
to a logarithmic dependence due to non-perturbative phenomena.}.
The limit of restoration of the $N=4$ supersymmetry, in both the theories,
corresponds to the decompactification of one radius ($R \to \infty$).
In this limit, also $U \to \infty$, and the effective coupling
constant, in the $N=4$ phase, depends only on the dilaton $S=S^{(\rm Het)}$,
as expected in both the theories\footnote{Still, there is the presence of
the logarithmic terms, both in $T$ and $U$. These terms can be lifted
by switching on an appropriate cut-off, as is discussed in
Refs.\cite{gkr,gkp,gkp2,solving}.}.
We consider the coincidence of the massless spectrum and the rather
non-trivial correspondence of these threshold corrections as compelling
evidence of the duality of the heterotic and the type I, model {\bf A}
constructions\footnote{In the construction with the
asymmetric shift in $\Gamma_{2,2}$ referred to in footnote 5,
there are lines, $T=f(U)$,
in the $(T,U)$ space, along which the ``smooth gravitational amplitude''
we are considering indeed becomes singular. This is due to the appearance
in the massless spectrum of new hypermultiplets, which are uncharged
under the gauge group of the torus, and therefore lead to
a jump in the $\beta$-function of the $R^2$ term, which is not
compensated by
an opposite jump in the $\beta$-function of $F^2$
(see Refs.\cite{gkp,kkprn}). By duality, these singularities should appear
also on the type I side, where new massless states should appear
for large or small values of the modulus $U$. The absence of these
rules out the alternative heterotic construction.}.

We now consider the limit $T \to 0$, with $U$ fixed.
In this limit, the theory
is better described in terms of the inverse modulus
$\tilde{T} \equiv -1/T$. By performing an $SL(2,Z)$ inversion,
the second term in Eq.(\ref{htr}) becomes
\be
-2 \log \Im \tilde{T} \left\vert \th_2 \left( \tilde{T}  \right)
\right\vert^4~,
\ee
which diverges linearly when $\Im \tilde{T}$ is large:
\be
\sim 2 \pi \Im \tilde{T} ~.
\ee
Therefore,
for large $\Im S^{(\rm Het)}$ and small $T$, Eq.(\ref{htr}) becomes
\be
{16 \, \pi^2 \over g^2_{\rm   grav}(\mu^{(\rm Het)})}
 \approx  \, 16 \, \pi^2 \Im S^{(\rm Het)}
+2 \pi \Im \tilde{T} -2 \Im U \vert \vartheta_4 \left( U  \right)\vert^4
 \, . \label{hthr2}
\ee
It is tantalizing to interpret the linear divergence in the field
$\tilde{T}$ as corresponding to the appearance of a D5-brane sector
in the dual type I theory.
Indeed, in this region of the moduli space, the effective coupling constant
of the heterotic theory behaves like that of the ``M-theory
breaking'' type I
model {\bf B}, Eq.(\ref{Ithrp}),
provided we identify the field $S^{(\rm Het)}$ with
$S$, as before\footnote{Here, by an abuse of language,
we are using the same notation, $S$, for both the type I constructions.}
and $\tilde{T}$ with $8 \pi S'=2\tau_S'$
(we have introduced here as usual the field $\tau_S'$, in analogy with
the field $\tau_S=4 \pi S$, the actual dilaton--axion field that
enters the $SL(2,Z)$, Montonen--Olive duality transformations).
As before, we can identify the complex structure moduli $U$
of the heterotic and type I constructions.
It therefore seems that, for large $\tilde{T}$, the heterotic
theory is perturbatively dual to the type I construction {\bf B}.
This correspondence is better understood in terms of the ``T-dual''
heterotic theory, obtained by exchanging the two radii of the torus,
$R_1 \leftrightarrow R_2$, and then
inverting them: $R_1 \to \tilde{R}_1= 1 / R_1$,
$R_2 \to \tilde{R}_2=1 / R_2$.
With these inversions, we remain in a spontaneously broken phase
of the $SO(16) \times SO(16)$ string,
but with the K\"{a}hler class modulus of the torus given
by $\tilde{T}$ instead of $T$.
The modulus $\tilde{T}$ is coupled
to the windings of the torus of this T-dual theory,
and in the limit of large $\Im \tilde{T}$ all the string states with
a non-zero winding number decouple from the spectrum, leaving only the
Kaluza--Klein states,  of the type I
dual string\footnote{This limit can also be viewed as the infinite-tension
limit of the heterotic string. We have in fact
$\Im \tilde{T} \sim \tilde{R}_1 \tilde{R}_2 / \alpha'$,
$\Im U \sim \tilde{R}_2 / \tilde{R}_1$,
and the limit $\Im \tilde{T} \to \infty$
with $U$ fixed is equivalent to the limit
$\alpha' \to 0$ with fixed radii.}.

The above duality implies in this example that the massless states of the
55-branes sector appear on the heterotic side as perturbative
states associated to the gauge currents, on the same footing as the
states of the 99-branes sector.
The situation is therefore rather different from that of
the models of Gimon and Polchinski \cite{gp}, in which the states
of the 55-branes sector are non-perturbative on the heterotic side
\cite{w,blpssw}. A solution to this (apparent) puzzle
comes from considering the T-dual, type I$^{\prime}$ picture \cite{adds},
in which the two gauge factors are provided by the
two ``Ho\v{r}ava--Witten'' walls of the M-theory, $S^1 \big/ Z_2$ orbifold,
where the D9-branes on one wall are wrapped on a T-dualized
four-torus, and effectively appear as D5-branes.
Nevertheless, because of their origin, the states corresponding
to open strings ending on these D-branes are expected to appear
as perturbative states on the heterotic theory\footnote{We thank
E. Dudas for a clarification of this point.}.

We can now try to see what happens from the type I side point of view
when the heterotic theory passes from small to large $T$
(or equivalently from large to small $\tilde{T}$).
Because of the identifications
\be
S^{(\rm Het)}=S={\rm e}^{-\phi_4}G^{1/4}\omega^{2}~,~~~~~~~~~~
\tilde{T}=8 \pi S'=8 \pi {\rm e}^{-\phi_4}G^{1/4}\omega^{-2}~,
\label{ssp}
\ee
we have $8 \pi S^{(\rm Het)} \big/ \tilde{T} = S \big/ S'= \omega^4$.
We therefore see that such a motion, performed while
keeping the field $S^{(\rm Het)}$ fixed, corresponds on the type I side
to increasing the volume of the $K3$, $\omega^4 \to \infty$.
In order to keep the field $S$ fixed, we have also to shrink
the volume of the two-torus and/or adjust the value of the dilaton $\phi_4$.
In order to remain in the phase of broken $N=4$ supersymmetry,
we must shrink the second circle, leaving the first one fixed.
Notice that, according to Eq.(\ref{vsvsp}), this motion is non-perturbative
from the type I point of view, involving a change of the
tree-level coupling constant.
In this limit, which corresponds to an effective
decompactification of the theory to eight dimensions
(or nine, if a circle of the two-torus is shrinked),
indeed the D5-branes look like D9-branes wrapped around the $T^2$ torus.
The $U(1)$ gauge bosons, which were provided by
open strings ending on the same D5-brane, are still there;
they contribute for a $U(1)^8$ factor to the gauge group, although
they must now be reinterpreted as due to strings ending on D9-branes.
What essentially distinguishes the behaviour of this model
with respect to the type I constructions, in which supersymmetry is
not spontaneously broken, as in  Refs.\cite{gp,gj},
is that there is an unbroken $S$-duality in those cases;
combined with the symmetry
under exchange of the fields $S$ and $S'$, i.e. of the 99- and 55-branes
sectors, this duality implies that,
along such a motion, non-perturbative phenomena enter
heavily in the game and, at the limit we are considering, the theory is
perturbatively described by the $S$-, $S'$-dual, identical theory.
In our case, instead, we expect $S$-duality to be broken,
because, as explained in Ref.\cite{adds},
the type I model {\bf B} indeed corresponds to a Scherk--Schwarz
mechanism applied to the 11-th dimension of
M-theory\footnote{We will come back to this point, which contradicts the
results of Ref.\cite{dg}.}.
It is then reasonable to find that in this limit,
since model {\bf B} is not falling back into itself, it ends up to
coincide with model {\bf A}, in which all the gauge bosons of $U(1)^{16}$
are provided by the D9-branes\footnote{In order to understand how
in this limit the D5-branes can look like D9-branes, consider the T-dual
situation in which the four circles of $T^4 \big/ Z_2$ are T-dualized.
In this case, for finite values of the radii, the D9-branes become
D5-branes and vice versa. However, when the radii are shrunk to zero,
the theory lives effectively in four extended and two compact dimensions,
where there are no 9-branes but only 5-branes wrapped around the compact
torus. T-dualizing again the four circles brings us back to
the original limit, in which all the D5-branes become D9-branes.}.

We now consider the restoration of the $N=4$ supersymmetry.
The higher amount of supersymmetry can be restored essentially
in two ways, which correspond on the type I side to the two
decompactifications: $R_2 \to \infty$ in model {\bf A} and
$R_1 \to 0$ in model {\bf B}.
In the first case, the restoration is perturbative in both
the type I and the heterotic side (the field $S \to \infty$).
In the second case, the restoration is non-perturbative from both
the type I and heterotic points of view ($S$, $S' \to 0$).
There are, however, also intermediate possibilities, which involve a
change  also of $\omega$. In these cases, the restoration,
although non-perturbative on the type I side, can look
perturbative on the heterotic side. This happens when
the product $G^{1/4} \omega^2$ is kept fixed.

\subsection*{\sl Non-perturbative corrections}

We have seen that through the duality between heterotic and type I
constructions, we gained insight into the non-perturbative behaviour of
both of them, at least regarding the restoration of the
$N=4$ supersymmetry.
We now try to go further: indeed, through the heterotic dual, we
learned that the two type I constructions are actually two
realizations of the same theory. The type I/heterotic
duality can be used to get insight into the non-perturbative corrections
to the effective coupling constant of the combination of
gravitational and gauge amplitudes given in Eq.(\ref{r2f2}).
 From the heterotic dual we know that the $SL(2,Z)_{S'}$ duality
group is broken to a $\Gamma(2)$ subgroup. On the heterotic side
the $\Gamma(2)_T \times \Gamma(2)_U$
group is by construction a symmetry that remains valid at any
value of the coupling. On the other hand,
we know that the type I ``M-theory breaking'' model {\bf B},
is perturbatively symmetric under the exchange of
the fields $S$ and $S'$: this is a consequence of the symmetry
under the exchange of the D9- and D5-branes sectors.
We claim that this implies that
also the  $SL(2,Z)_{S}$ duality group is indeed broken in the same way
as the $SL(2,Z)_{S'}$ group.
This statement, which promotes a perturbative symmetry
to a non-perturbative one, is supported by the observation that,
as is discussed in Ref.\cite{adds},
in the T-dual, type I$^{\prime}$ picture, the two contributions come
from the
two ``Ho\v{r}ava--Witten'' walls of the M-theory on $S^1 \big/ Z_2$,
and the symmetry of the problem under exchange of the two
remains true at any value of the 11-th coordinate.
Once observed that the symmetry of the theory is
$\Gamma(2)_{2S} \times \Gamma(2)_{2S'}\times \Gamma(2)_U$ times the
permutations of the three factors, we can write
the full, non-perturbative correction, which reduces to  
Eq.(\ref{htr}) in the
large-$S$ limit:
\ba
{16 \, \pi^2 \over g^2_{\rm   grav}(\mu)}
& = &
-2 \log \Im \tau_S
\left \vert \vartheta_2 \left( 2 \tau_S \right) \right\vert^4
-2 \log \Im \tau_S'
\left \vert \vartheta_2 \left( 2 \tau_S' \right) \right\vert^4
-2 \log \Im U \left \vert \vartheta_4 \left( U \right) \right\vert^4
\nn \\
&& +{\cal E}(2\tau_S,2\tau_S',U)+ 14 \log {M \over \mu }~.
\label{tr}
\ea
In this expression we used the type I fields $\tau_S=4 \pi S$,
$\tau_S'=4 \pi S'$ and $U$,
which by now we know to be the same for both the type I models
and equivalent to the heterotic fields $\tau_S^{\rm Het}=4 \pi S^{\rm Het}$,
$\tilde{T}/2$ and $U$.
The infrared cut-off $\mu$ and the mass scale $M$
can be indifferently those of the type I or of the heterotic string,
their relation being
\be
{ M^{(\rm Het)} \over \mu^{(\rm Het)}} =
{ M^{(\rm I)} \over \mu^{(\rm I)}}~.
\ee
In Eq.(\ref{tr}) we allow for the presence of a series,
${\cal E}(2\tau_S,2\tau_S',U)$, of exponentials symmetric in
$2\tau_S, 2\tau_S', -1/U$.
Such a term, always suppressed in the perturbative limit,
cannot be excluded by the symmetries of the theory.

In each of the limits in which the $N=4$ supersymmetry
is perturbatively restored (either on the heterotic or on both sides),
the contribution of two moduli drops out
(it reduces to the already mentioned logarithmic dependence)
and the correction Eq.(\ref{tr}) diverges linearly as a function of only one
modulus (the field $S$, as expected for $N=4$ corrections).
The term ${\cal E}(2\tau_S,2\tau_S',U)$
is suppressed in the limits of a restoration
of supersymmetry. By using its symmetry properties it is easy to see that
it is suppressed also in the non-perturbative limit $S \to 0$.
The form of Eq.(\ref{tr}) therefore tells us that there exists a limit,
which is non-perturbative on both the heterotic and type I sides,
in which there is an effective restoration of an $N=8$ supersymmetry.
This takes place  when the three moduli contributions drop out,
namely when $S \to 0$, $S' \to 0$ and $U \to \infty$.
This for instance happens when, in model {\bf B}, we
send $R_1$ to zero while keeping $\omega$, $R_2$
and the field $\phi_4$ fixed. In this case the effective coupling  
Eq.(\ref{tr})
vanishes (modulo the usual logarithmic divergences).

Through the above analysis we have learned
that in this theory, which can be interpreted
as obtained by a freely acting projection of M-theory,
the $U$-duality group
$SL(2,Z)_S \times SL(2,Z)_T \times SL(2,Z)_U$ is indeed broken.
In particular, the $S$-duality is broken to a $\Gamma(2)_{2\tau_S}$
subgroup. This is in contrast with the results of Ref.\cite{dg},
in which, even in the presence of a Scherk--Schwarz
compactification of the 11-th coordinate of M-theory, $S$-duality
remains unbroken.
We believe that, once properly treated, a freely acting projection
applied on the 11-th coordinate must instead necessarily break the
$S$-duality.
The result of Ref.\cite{dg} is obtained by using as a starting point
the effective Ho\v{r}ava--Witten action, in which the space-time
metric appearing in the gauge terms,
which live on the two ten-dimensional walls, is a ``coordinate-independent''
restriction of the eleven-dimensional metric to ten dimensions.
In this way, the heterotic dilaton has no dependence
on the 11-th radius.
This scenario is correct for a genuine Ho\v{r}ava--Witten
orbifold of M-theory, in which the original amount of supersymmetry
can never
be restored by a decompactification of the eleventh dimension.
This is, however, not the case of a Scherk--Schwarz
compactification of M-theory,
in which we expect the supersymmetry to be restored in the large-radius
limit (as happens in our present case, where the radius of M-theory
is T-dual to the $R_1$ of model {\bf B}). In this limit, the gauge bosons
of the terms introduced to cancel the ten-dimensional anomaly
should decouple. This is indeed what happens in our case,
and is consistent with general arguments that fix
the relation between the four-dimensional couplings and the eleventh radius
$\rho$ to be \cite{hw,aq}:
\be
\Im S,\, \Im S' \, \sim \, \rho^{-2 / 3}~,
\ee
where we have omitted the factors that contain all the other parameters.

As a last point, we remark
that the model is symmetric under permutations of the fields
$X=8 \pi S$, $Y=8 \pi S'$  and $-1/ U$.
By proceeding as in Ref.\cite{gkp2},
it is therefore possible to calculate (at least a part of)
the non-perturbative prepotential.
This is obtained by a proper symmetrization of the perturbative
prepotential, computed, as a function of the moduli $T$ and $U$,
on the heterotic side.
The result is the same as in Ref.\cite{gkp2}.

\noindent

\vskip 0.3cm
\setcounter{section}{5}
\setcounter{equation}{0}
\section*{\normalsize{\bf 5. Discussion}}

In the previous section we were able to determine
part of the non-perturbative behaviour of the gravitational corrections.
In order to determine also the term ${\cal E}(2\tau_S,2\tau_S',U)$
in Eq.(\ref{tr}), we would need to identify a type IIA dual,
in which these three fields would belong to the perturbative moduli.
This would allow us to repeat the analysis of  
Refs.\cite{gkp,hmn=4,gkp2,hmn=2}.
In order for the duality to work, such a type IIA dual would have to be
constructed as a compactification on (an orbifold limit of)
a K3 fibration \cite{kv}--\cite{al}.
A necessary condition is therefore a symmetric breaking of the
supersymmetry;
this requirement leads directly to the type IIA orbifold of Section 2.
However, the heterotic and the type IIA orbifolds under consideration
cannot be compared for finite values of the heterotic
moduli $T$ and $U$.
This is related to the fact that the compact space of the type IIA orbifold,
a $T^6/( Z_2 \times Z_2)$ limit of $CY^{19,19}$, cannot be seen
as a singular limit of a K3 fibration (see Refs.\cite{vm}--\cite{gkr}).
Indeed, the spaces of the vector moduli of the two models
do not correspond.
A signal of the non-coincidence of these subspaces is provided by the
absence of a perturbative super-Higgs phenomenon on the type IIA
construction: unlike in the heterotic model, on the type IIA side
it is not possible to restore the $N=4$ supersymmetry in a corner of
the space
of moduli $T^i$, $U^i$, $i=1,2,3$
\footnote{As is discussed in Ref.\cite{gkr},
compactification on a K3 fibration and spontaneous breaking of
the $N=4$ supersymmetry are directly related.}.
However, these models can be viewed as different phases of a
single theory, corresponding to different regions
of a wider moduli space\footnote{We consider here only the space of moduli
belonging to the vector multiplets.}.
These two regions are connected at the
limits $\Im T^2 \to 0,\infty$, $\Im T^3 \to 0,\infty$ of the type IIA
moduli space, for any value of the modulus $T^1$.
In these limits, the type II correction, Eq.(\ref{IIthr}),
reproduces the heterotic/type I, Eq.(\ref{tr}), at the
appropriate corners in the space of moduli $T$ and $U$
($\Im T, \Im U \to 0$ and/or $\infty$).
For any value of $T^1$, we have the identifications
$T^1=\tau_S^{\rm As}=-1 \big/ 2\tau_S$.
Indeed, for any value of this modulus, there are two branches of the theory:
a branch corresponds to a heterotic/type I phase, with moduli $T,U$;
crossing the borders at $\Im T,\Im U \to 0,\infty$, one passes to
another branch, which corresponds to the type II phase,
described by the type IIA or type II asymmetric orbifolds,
with moduli $T^2$, $T^3$.
We can gain a deeper understanding of the reason why the heterotic
and the type IIA constructions cannot be dual, if we consider
the relation between these heterotic/type I constructions
and the heterotic/type IIA, $S,T,U$ models with $N_V=H_H=0$ of  
Ref.\cite{gkp2},
which also possess a spontaneously broken $N=8$ supersymmetry.
In that case, the spectrum is simply the truncation of the $N=8$
supergravity, and the restoration of the $N=8$ supersymmetry,
which is non-perturbative from the heterotic point of view,
being related to a motion in the dilaton field $S$,
necessarily appears as perturbative on the type IIA side,
where the field $S$ is the volume form of the K3 fibration.
Indeed, on the type IIA side, the spontaneous breaking of the $N=8$
supersymmetry is due to a perturbative Scherk--Schwarz mechanism
realized through freely acting projections, and is therefore
directly related to the absence of extra massless states coming from
orbifold
twisted sectors. On the other hand,
the heterotic/type I theory considered in this paper
is an interesting example of (non-perturbative)
spontaneous breaking of the $N=8$ supersymmetry,
which, possessing an enlarged gauge group,
cannot correspond to a truncation of the type IIA, $N=8$ spectrum,
but is directly related to another phase of the M-theory.
The two phases are interpolated by switching on and turning off
``Wilson lines'', which act as freely acting $Z_2$ projections
twisting all the gauge bosons of the two ``Ho\v{r}ava--Witten'' planes.
As is clear from the partition function given in Eqs. (4.2)--(4.8)
of Ref.\cite{gkp2}, they appear on the heterotic side
as ``$N=4$'' Wilson lines $Y$, which
act on the twisted coordinates, corresponding to the hypermultiplets.
Switching on/off these Wilson lines involves a blowing up of the
moduli frozen at the fixed points of $T^4 \big/ Z_2$ and a motion to another
$T^4 \big/ Z_2$ singularity\footnote{This necessarily involves passing
through a $N=4$ phase of the model, so that the results of Ref.\cite{gkp2}
for the $R^2$ correction cannot be used to rule out the term
${\cal E}(2\tau_S,2\tau_S',U)$ in Eq.(\ref{tr}).}.
Such a motion is on the other hand non-perturbative
from the type IIA point of view, the moduli associated to $Y$
including the dual of the type II dilaton field.

The type II constructions of Section 2, on the other hand,
correspond to a limit in the moduli space in which the
$N=8$ supersymmetry is not spontaneously broken; in this limit,
according to Ref.\cite{adds}, on the type I$^{\prime}$ picture,
the combined action of the $Z_2$ which acts as a translation
on the 11-th coordinate of M-theory, breaking $N=8$ to $N=4$, and
that of the $Z_2$ which further breaks to $N=2$, is no longer free.
{}From a perturbative point of view, the region of the moduli space
corresponding to this limit is achieved at the above specified corners
in the space of moduli $T^2, T^3$ (or $T^{(\rm Het)},U^{(\rm Het)}$), where
the two theories can be connected.
The above discussion is sketched in Fig. 1.\\
\begin{figure}[htb]
  \begin{center}
  \makebox[8cm]{
     \epsfxsize=14cm
     \epsfysize=10cm
     \epsfbox{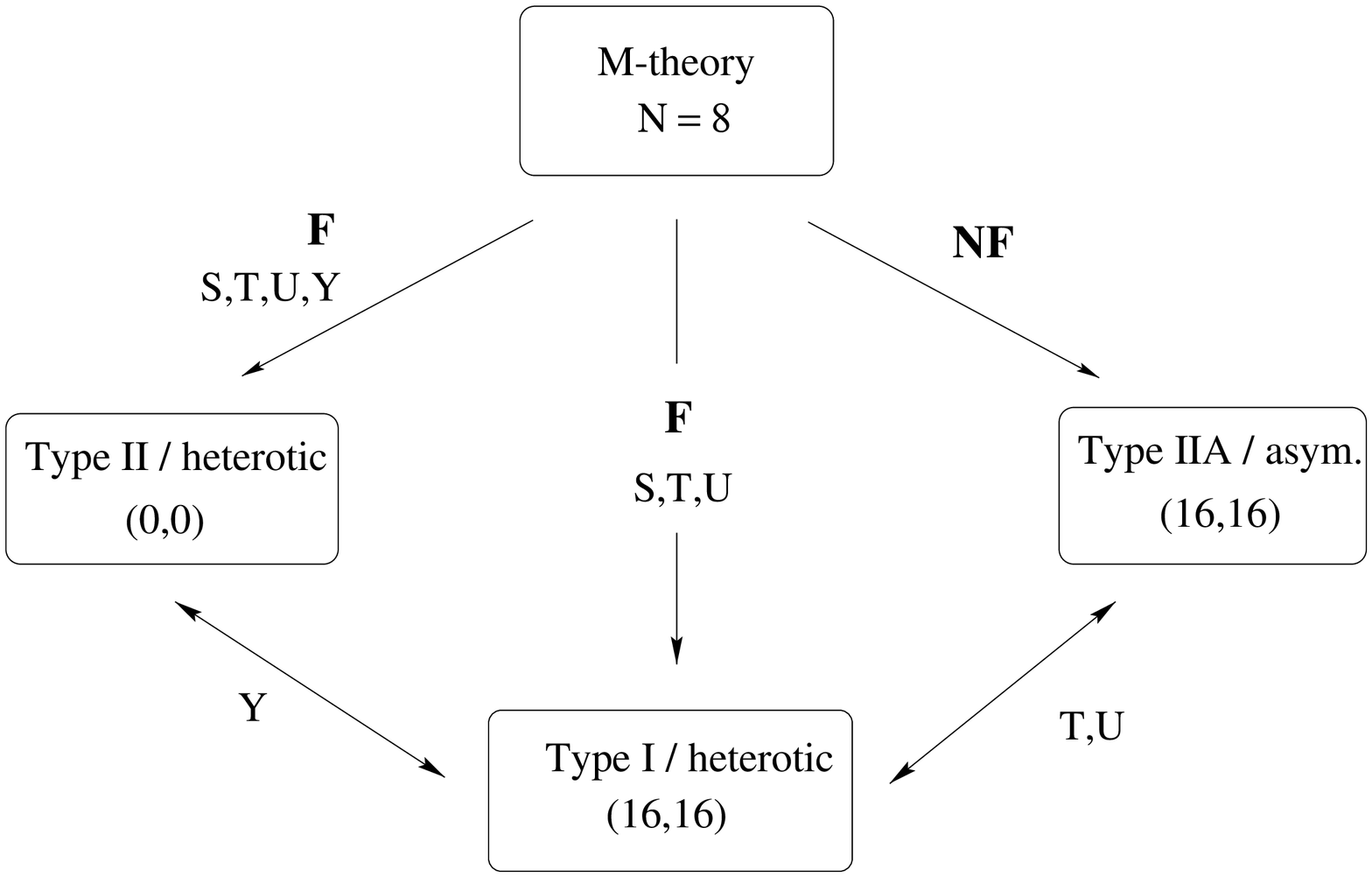}
  }
  \end{center}
  \vspace{1.4cm}
  \refstepcounter{figure}
  \parbox{15cm}{\hspace{1.55cm}
      \parbox{13cm}{
           Figure \thefigure : The connections between the $N=2$ models
           and M-theory.
           ({\bf N}){\bf F} indicates a (non-)freely acting orbifold
           projection.
           With $(N_V,N_H)$ we indicate the number of vector and
           hypermultiplets of the ``twisted sector''.
      }
  }
  \label{figuretype1}
\end{figure}

\vskip 0.3cm
\setcounter{section}{6}
\setcounter{equation}{0}
\section*{\normalsize{\bf 6. Conclusions}}

In this paper we investigated the connections between several
four-dimensional string
constructions with the same massless spectrum,
namely an $N=2$ supergravity with $3+N_V$ vector multiplets
and $4+N_H$ hypermultiplets, and $N_V=N_H=16$.
This spectrum is obtained via type IIA/B, heterotic and type I
orbifold compactifications.
We found evidence that the two type I constructions
with spontaneous breaking of supersymmetry,
presented in Ref.\cite{adds}, namely the $N=2$ ``Scherk--Schwarz''
and ``M-theory'' breaking models, indeed constitute two
phases of the same theory, and are non-perturbatively related
by a motion in the field $S'$, which parametrizes the coupling
constant of the gauge fields of the D5-branes sector.
This relation appears as perturbative on the heterotic dual construction.
Collecting the knowledge coming from the heterotic model and the type
I duals,
we got some insight into the non-perturbative aspects of this theory.
In particular, we discovered the existence of a non-perturbative
super-Higgs phenomenon responsible for the spontaneous breaking of
the $N=8$ supersymmetry. This is consistent with the interpretation
of the theory as due to a ``Scherk--Schwarz'' mechanism applied to the 11-th
dimension of the M-theory.
This mechanism is also responsible for the breaking
of the $SL(2,Z)$, Montonen--Olive $S$-duality, to a $\Gamma(2)$ subgroup,
and it reflects on the dilaton dependence of string corrections
to effective coupling constants, as the gravitational ones
we considered.
On the other hand, these constructions do not possess type II duals.
Indeed we show that the type IIA, type II asymmetric orbifolds
with the same massless spectrum actually correspond
to a different phase of the M-theory. The two phases are
connected at certain corners in the moduli space.

\vskip 1.cm
\centerline{\bf Acknowledgements}
\noindent
We thank R. Blumenhagen, E. Dudas, B. K\"{o}rs, A. Miemiec, H. Partouche,
P.M. Petropoulos, A. Sagnotti and D. Smith for valuable discussions.

This work was partially supported by the EEC under the contract
TMR-ERBFMRX-CT96-0045.

\end{document}